\pdfoutput=1
\documentclass{ptephy_v1}

\usepackage{amsmath,amsthm}
\usepackage{color}

\newcommand{\sgn}{\mathrm{sgn}}

\newcommand{\Det}{\mathrm{Det}}

\begin{document}

\title{One-dimensional anyons in relativistic field theory}
\author{Arata~Yamamoto}
\affil{Department of Physics, The University of Tokyo, Tokyo 113-0033, Japan \email{arayamamoto@nt.phys.s.u-tokyo.ac.jp}}

\begin{abstract}
We study relativistic anyon field theory in 1+1 dimensions.
While (2+1)-dimensional anyon fields are equivalent to boson or fermion fields coupled with the Chern-Simons gauge fields, (1+1)-dimensional anyon fields are equivalent to boson or fermion fields with many-body interaction.
We derive the path integral representation and perform the lattice Monte Carlo simulation.
\end{abstract}

\subjectindex{B38}

\maketitle

\section{Introduction}

The exotic particle obeying fractional statistics or interpolating a boson and a fermion is called the anyon \cite{1977NCimB..37....1L,1982PhRvL..49..957W,Arovas:1985yb}.
Although the anyon does not exist as an elementary particle in a vacuum, it can arise as a collective excitation in a matter.
The most famous experimental manifestation is the fractional quantum Hall effect \cite{2008AnPhy.323..204S}.
The anyon also appears as Majorana zero modes on topological superconductors.
The topological nature of the anyon is hopeful for the application to quantum computation \cite{RevModPhys.80.1083}.

The anyon was originally proposed in 2+1 dimensions.
The field theoretical description of the (2+1)-dimensional anyon has been well understood.
It is given by ordinary boson or fermion fields coupled with the Chern-Simons gauge fields.
Even if non-interacting anyon field theory is considered, it is equivalent to interacting gauge theory with a variety of quantum phenomena.
From a practical point of view, the Monte Carlo simulation of the lattice Chern-Simons gauge theory is difficult due to sign problem, doubling problem, and gauge symmetry breaking \cite{Frohlich:1988qh,Luscher:1989kk,Muller:1990xd,Kantor:1991ty,Eliezer:1990iy,Eliezer:1991cv,Eliezer:1992sq,Eliezer:1991qh,Diamantini:1993iu,Adams:1997eb,Berruto:2000dp,Bietenholz:2000ca,Fosco:2001aq,Nagata:2008zza,MacKenzie:2010xk}.

After the anyon was proposed in 2+1 dimensions, it was generalized to arbitrary dimensions \cite{1991PhRvL..67..937H}.
Since anyons in other dimensions are novel quantum states, their physical properties are fascinating subjects.
In particular, (1+1)-dimensional anyons and its experimental realization have been intensively discussed in non-relativistic physics \cite{1999PhRvL..83.1275K,2006PhRvL..96u0402B,2007PhRvB..75w3104C,2008PhRvA..78b3631H,2009PhRvA..79d3633H,2011NatCo...2E.361K,2012PhRvA..86d3631H,2014PhRvL.113e0601W,2014PhRvA..90f3618W,2015PhRvL.115e3002G,2015NJPh...17l3016T,2015PhRvA..92f3634Z,2016PhRvL.117t5303S,2016PhRvA..93f3627H,2016PhRvA..94a3611A,2016JSMTE..07.3106M,2017PhRvA..95f3621L}.
On the other hand, the study of (1+1)-dimensional relativistic anyons has been limited in quantum mechanics \cite{Gamboa:1995np,Gamboa:1996xh}.
In particle physics or for the condensed matters with relativistic dispersion, the extension to relativistic field theory would be interesting.

In this paper, we study the relativistic version of anyon field theory in 1+1 dimensions.
While the quantum anyon field in the operator formalism is well-defined, the classical (single-valued) anyon field in the path integral formalism is ill-defined.
We start with the Hamiltonian operator of the anyon field, transform the anyon field to the boson field, and then derive the path integral representation.
We also present the first attempt of the lattice Monte Carlo simulation.
The simulation of the (1+1)-dimensional anyon field theory is much easier than the (2+1)-dimensional one, even though it has the sign problem, as shown in this paper.

\section{Commutation relation}

In 1+1 dimensions, the anyon field $\Phi$ and the conjugate field $\Pi$ satisfy the commutation relation
\begin{equation}
\label{eqCCA}
\begin{split}
 \Phi(x) \Phi(y) &= e^{i\theta \sgn(x-y)} \Phi(y) \Phi(x)
\\
 \Pi(x) \Pi(y) &= e^{-i\theta\sgn(x-y)} \Pi(y) \Pi(x)
\\
 \Phi(x) \Pi(y) &= e^{-i\theta \sgn(x-y)} \Pi(y) \Phi(x) + i\delta(x-y)
.
\end{split}
\end{equation}
The sign function is defined by
\begin{equation}
\label{eqSF}
 \sgn(x-y) = 
  \begin{cases}
    +1 \quad &(x>y)\\
    0  \quad &(x=y)\\
    -1 \quad &(x<y)
  \end{cases}
.
\end{equation}
The sign function explicitly breaks the space-inversion symmetry $x \to -x$.
One anyon gets the phase $+\theta$ when going through the other anyon in the $+x$ direction, and gets the phase $-\theta$ when coming back in the $-x$ direction.
Thus the total circulation is always zero.
This is contrast with nonzero circulation of (2+1)-dimensional anyons, which are defined by the commutation relation $\Phi(x) \Phi(y) = e^{i\theta} \Phi(y) \Phi(x)$ etc.

We introduce a complex scalar field
\begin{equation}
\phi(x) = \frac{1}{\sqrt{2}} (\phi_1(x) + i \phi_2(x))
,
\end{equation}
the conjugate field
\begin{equation}
\pi(x) = \frac{1}{\sqrt{2}} (\pi_1(x) - i \pi_2(x))
,
\end{equation}
and the number density operator
\begin{equation}
\begin{split}
\label{eqN}
 n(x) &= i \pi(x) \phi(x) - i \pi^*(x) \phi^*(x)
\\
&= \pi_2(x) \phi_1(x) - \pi_1(x) \phi_2(x)
.
\end{split}
\end{equation}
These operators satisfy the bosonic commutation relation
\begin{equation}
\begin{split}
\label{eqCCB}
[\phi(x),\pi(y)] &= i\delta(x-y) \\ 
[ n(x), \phi(y) ] &= \phi(x)\delta(x-y) \\
[ n(x), \pi(y) ] &= - \pi(x)\delta(x-y)
.
\end{split}
\end{equation}
The anyonic commutation relation \eqref{eqCCA} can be realized by the Jordan-Wigner transformation
\begin{equation}
\label{eqJW}
\begin{split}
 \Phi(x) &= \phi(x) \exp \left( i\theta \int_{-\infty}^{x} n(z) dz \right)
\\
 \Pi(x) &= \exp \left( - i\theta \int_{-\infty}^{x} n(z) dz \right) \pi(x) 
.
\end{split}
\end{equation}
We can easily show that Eq.~\eqref{eqJW} satisfies Eq.~\eqref{eqCCA}.
From Eq.~\eqref{eqJW}, a single anyon can be interpreted as a composite boson attached with the line of number density.
This is analogous to the picture of a (2+1)-dimensional anyon as an ordinary particle attached with the vortex of the Chern-Simons gauge field.

The above construction is the generalization of non-relativistic anyons in 1+1 dimensions \cite{1999PhRvL..83.1275K}.
The difference is that the conjugate field must be explicitly introduced in relativistic theory.
This results in the complexity of the path integral representation, as shown in the next section.
(In non-relativistic theory, the conjugate field can be formally introduced but it is just an auxiliary field.
Therefore the path integral representation is trivial.)

\section{Path integral}

We derive the path integral representation of anyon field theory.
We show the simplified derivation of the path integral.
Some intermediate steps, e.g., inserting the coherent states, are assumed and omitted.
The suitable ordering of operators to make classical and quantum equations coincident is also assumed.

Let us consider the Hamiltonian
\begin{equation}
\label{eqHA}
\mathcal{H} = \Pi^* \Pi + \partial_1 \Phi^* \partial_1 \Phi + V
,
\end{equation}
where $V$ is a local potential term.
By the Jordan-Wigner transformation \eqref{eqJW}, the Hamiltonian is rewritten by the scalar fields as
\begin{equation}
\label{eqHB}
\mathcal{H} = \pi^* \pi + (\partial_1 - i\theta n) \phi^* (\partial_1 + i\theta n) \phi  + V
.
\end{equation}
The second term, the kinetic term, looks like the covariant derivative with $n$, but it does not have gauge covariance like the Chern-Simons gauge theory \cite{PhysRevLett.76.4007,PhysRevLett.77.4851,Aglietti:1996mv}.
If we take $n$ as a mean field, its effect is just a constant shift of momentum $p \to p +\theta n$.
Thus, in the mean-field approximation, the anyon field theory \eqref{eqHA}, except for the shift of momentum, has the same property as scalar field theory.
Beyond the mean-field approximation, the second term includes nontrivial many-body interaction.

The Lagrangian is obtained by the Legendre transformation
\begin{equation}
\mathcal{L} = \pi_i \partial_0 \phi_i - \mathcal{H} + \mu n
.
\end{equation}
Here $\mu$ is a chemical potential.
Following the conventional derivation, we complete the square
\begin{equation}
\begin{split}
\mathcal{L}
&= - \frac{1}{2} \pi_i A_{ij} \pi_j + \pi_i B_i - \frac{1}{2} (\partial_1 \phi_i)^2 - V
\\
&= - \frac{1}{2} (\pi_i - B_k A_{ki}^{-1}) A_{ij} (\pi_j - A_{jl}^{-1}B_l) + \frac{1}{2} B_i A_{ij}^{-1} B_j - \frac{1}{2} (\partial_1 \phi_i)^2 - V
\end{split}
\end{equation}
with
\begin{eqnarray}
A &=&
 \begin{pmatrix}
 1 + \theta^2 \phi_2^2 \phi_i^2 & - \theta^2 \phi_1 \phi_2 \phi_i^2
 \\
 - \theta^2 \phi_1 \phi_2 \phi_i^2 & 1 + \theta^2 \phi_1^2 \phi_i^2
 \end{pmatrix}
\\
B_1 &=& \partial_0 \phi_1 - \nu \phi_2
\\
B_2 &=& \partial_0 \phi_2 + \nu \phi_1
\\
\nu &=& \mu - \theta (\phi_1 \partial_1 \phi_2 - \phi_2 \partial_1 \phi_1)
.
\end{eqnarray}
The repeated indices, $i,j,k,l$, are summed over, e.g., $\phi_i^2 = \phi_1^2 + \phi_2^2$.
Performing the Gaussian integration of $\pi_i$, we obtain the path integral
\begin{equation}
\label{eqZ}
\begin{split}
 Z 
&= \int D\phi_1 D\phi_2 \frac{1}{\sqrt{\Det A}} e^{i\int d^2 x \mathcal{L}}
\\
&= \int D\phi_1 D\phi_2 \frac{1}{\sqrt{\Det \left( 1 + \theta^2 (\phi_k^2)^2 \right)}} e^{i\int d^2 x \mathcal{L}}
\end{split}
\end{equation}
with the Lagrangian
\begin{equation}
\label{eqLB}
\begin{split}
\mathcal{L}
&= \frac{1}{2} B_i A_{ij}^{-1} B_j - \frac{1}{2} (\partial_1 \phi_i)^2 - V
\\
&= \frac{1}{2}\frac{1}{1 + \theta^2 (\phi_k^2)^2} \left\{ (\partial_0 \phi_i)^2 + \theta^2 \phi_i^2 (\phi_j \partial_0 \phi_j)^2 + 2\nu ( \phi_1 \partial_0 \phi_2 - \phi_2 \partial_0 \phi_1 ) + \nu^2 \phi_i^2 \right\}
\\
&\quad - \frac{1}{2} (\partial_1 \phi_i)^2 - V
.
\end{split}
\end{equation}
Since the Hamiltonian \eqref{eqHB} depends on the conjugate field in a nontrivial manner, we obtained the complicated form of the path integral.
At $\mu=0$, the Lagrangian is
\begin{equation}
\begin{split}
\mathcal{L}
&= \frac{1}{2}\frac{1}{1 + \theta^2 (\phi_k^2)^2} \big\{ (\partial_0 \phi_i)^2 + \theta^2 \phi_i^2 (\phi_j \partial_0 \phi_j)^2 - (\partial_1 \phi_i)^2 - \theta^2 \phi_i^2 (\phi_j \partial_1 \phi_j)^2
\\
&\quad - 2 \theta ( \phi_1 \partial_0 \phi_2 - \phi_2 \partial_0 \phi_1 ) (\phi_1 \partial_1 \phi_2 - \phi_2 \partial_1 \phi_1) \big\} - V
.
\end{split}
\end{equation}
The Lagrangian, except for the $O(\theta)$ term, is Lorentz symmetric.
The $O(\theta)$ term only preserves the simultaneous inversion symmetry, $t\to -t$ and $x\to -x$, and breaks each inversion symmetry, $t\to -t$ or $x\to -x$.
This originates from inversion symmetry breaking of the commutation relation \eqref{eqCCA}.

The correlation functions of the anyon field are rewritten by the correlation functions of the scalar field.
For example, the anyon number density is equal to the boson number density because $n = i \Pi \Phi - i \Pi^* \Phi^* = i \pi \phi - i \pi^* \phi^*$.
Inserting
\begin{equation}
\begin{split}
\pi_1
&= \frac{\partial \mathcal{L}}{\partial (\partial_0 \phi_1)}
= \frac{1}{1 + \theta^2 (\phi_k^2)^2} \big\{ \partial_0 \phi_1 + \theta^2 \phi_i^2 \phi_1 (\phi_j \partial_0 \phi_j) - \nu \phi_2 \big\}
\\
\pi_2
&=\frac{\partial \mathcal{L}}{\partial (\partial_0 \phi_2)}
= \frac{1}{1 + \theta^2 (\phi_k^2)^2} \big\{ \partial_0 \phi_2 + \theta^2 \phi_i^2 \phi_2 (\phi_j \partial_0 \phi_j) + \nu \phi_1 \big\}
,
\end{split}
\end{equation}
we obtain the path integral representation of the number density
\begin{equation}
\label{eqn}
n = \frac{1}{1 + \theta^2 (\phi_k^2)^2} \big( \phi_1 \partial_0 \phi_2 - \phi_2 \partial_0 \phi_1 + \nu \phi_i^2 \big)
.
\end{equation}
With these expressions, we can check the consistency between the Hamiltonian and the Lagrangian.
For example, the number density \eqref{eqn} is consistent with the one derived from the Noether theorem.

The anyonic commutation relation \eqref{eqCCA} has the $2\pi$-periodicity of $\theta$.
The Hamiltonian \eqref{eqHB} preserves the $2\pi$-periodicity.
This can be checked by considering the lattice Hamiltonian.
The $\theta$-dependent term is discretized as $(\partial_1-i\theta n)\phi^*(\partial_1-i\theta n)\phi \to \{e^{-i\theta na} \phi^*(x+a) - \phi^*(x)\}\{e^{i\theta na} \phi(x+a) - \phi(x)\}/a^2$.
In the particle number basis, this is manifestly $2\pi$-periodic.
In principle, since the grand canonical ensemble is a superposition of canonical ensembles and each canonical ensemble is $2\pi$-periodic, the path integral representation must be also $2\pi$-periodic.
It is, however, unclear in the ground canonical Lagrangian \eqref{eqLB}.
This is because particle numbers are not integers in the coherent state basis.
This is different from (2+1)-dimensional anyons, of which the $2\pi$-periodicity can be proved by the topological quantization of the Chern-Simons gauge field.
In 1+1 dimensions, the $2\pi$-periodicity originates from particle number quantization, not from topological quantization.
It is not easily seen in the path integral formalism.

\section{Monte Carlo simulation}

We performed the Monte Carlo simulation of this path integral.
Although the path integral representation is complicated, it is not so serious for the Monte Carlo simulation.
By the Wick rotation $\partial_0 \to i\partial_2$, we obtain the Euclidean Lagrangian
\begin{equation}
\begin{split}
\mathcal{L}_E &= {\rm Re} \mathcal{L}_E + i {\rm Im} \mathcal{L}_E
\\
{\rm Re} \mathcal{L}_E
&= \frac{1}{2} \frac{1}{1 + \theta^2 (\phi_k^2)^2} \big\{ (\partial_2 \phi_i)^2 + \theta^2 \phi_i^2 (\phi_j \partial_2 \phi_j)^2 - \nu^2 \phi_i^2 \big\} + \frac{1}{2} (\partial_1 \phi_i)^2 + V
\\
{\rm Im} \mathcal{L}_E &= - \frac{1}{1 + \theta^2 (\phi_k^2)^2} \nu ( \phi_1 \partial_2 \phi_2 - \phi_2 \partial_2 \phi_1 ) 
\end{split}
\end{equation}
Note that the Euclidean Lagrangian is complex and thus has the sign problem.
We introduced the quartic interaction potential $V = (g/4) (\phi_i^2)^2$ for the absolute value of $\phi_i$ not to blow up.
We discretized the Euclidean Lagrangian with lattice spacing $a$ and performed the Hybrid Monte Carlo simulation of lattice scalar field theory \cite{Montvay:1994cy}.
Although there is the sign problem, it can be solved by the brute-force reweighting method because the computation of two-dimensional lattice scalar field theory is cheap.
The quartic coupling constant $g$ and the temperature $T$ are fixed at $ga^2=1$ and $T a = 0.1$.
The temporal boundary condition was periodic and the spatial boundary condition was Neumann.

\begin{figure}[h]
\begin{center}
 \includegraphics[width=.7\textwidth]{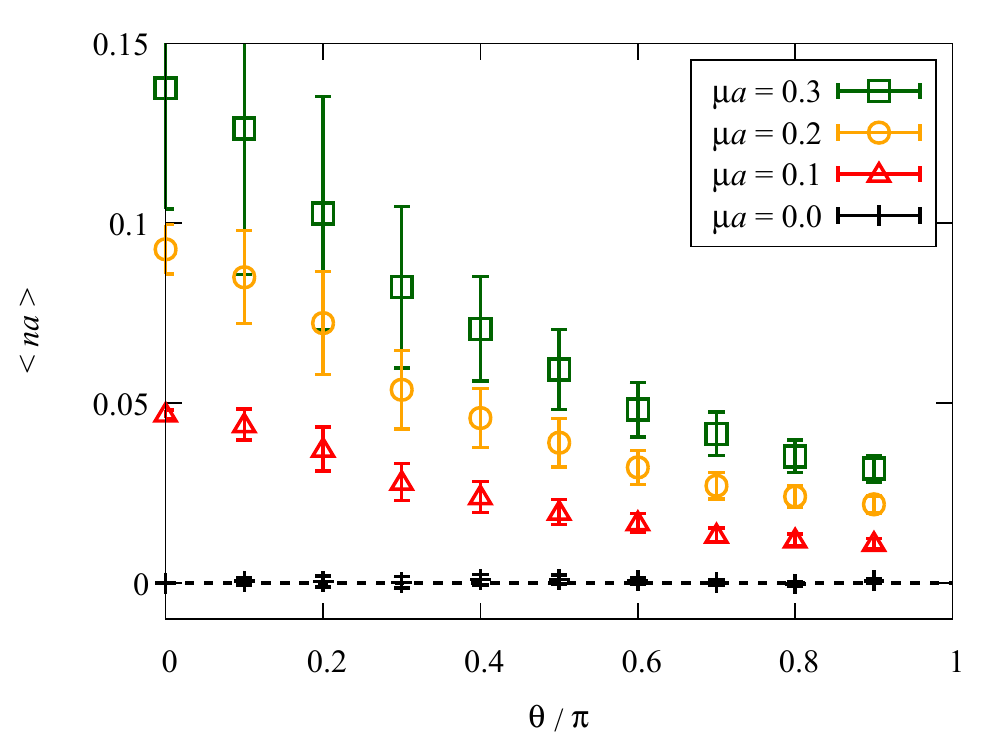}
\caption{
Number density $\langle n \rangle$ as a function of the statistical parameter $\theta$.
The spatial volume is fixed at $L/a =10$.
\label{figM}
}
\end{center}
\end{figure}

\begin{figure}[h]
\begin{center}
 \includegraphics[width=.7\textwidth]{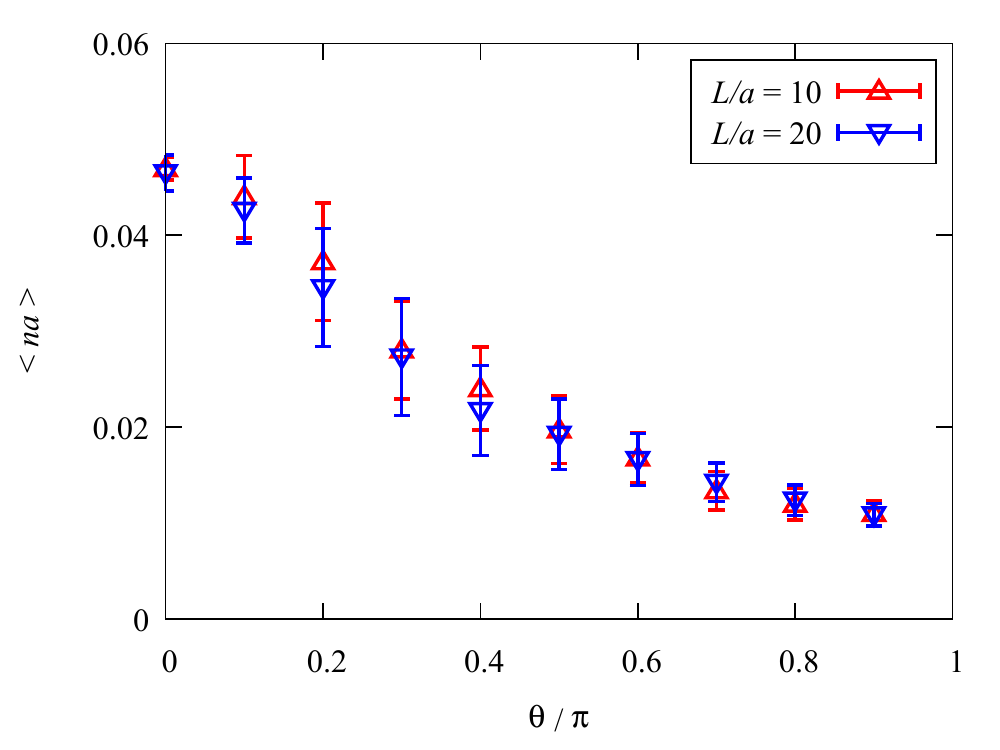}
\caption{
Number density $\langle n \rangle$ as a function of the statistical parameter $\theta$.
The chemical potential is fixed at $\mu a =0.1$.
\label{figN}
}
\end{center}
\end{figure}

We calculated the averaged number density, $n \equiv (T/L)\times \int d^2x \, n(x,\tau)$, where $L$ is the spatial volume.
In the reweighting method, the expectation value of the number density is given by
\begin{equation}
 \langle n \rangle = \frac{\langle n e^{-i\int d^2x {\rm Im}\mathcal{L}_E} \rangle_R}{\langle e^{-i\int d^2x {\rm Im}\mathcal{L}_E} \rangle_R}
,
\end{equation}
where $\langle \cdots \rangle_R$ is the expectation value in the real part of the path integral
\begin{equation}
 Z = \int D\phi_1 D\phi_2 \frac{1}{\sqrt{\Det \left( 1 + \theta^2 (\phi_k^2)^2 \right)}} e^{-\int d^2 x {\rm Re} \mathcal{L}_E}
.
\end{equation}
The simulation results are shown in Fig.~\ref{figM}.
The number density is a decreasing function of the statistical parameter $\theta$.
The decreasing function means that the statistical parameter promotes repulsive interaction.
This can be understood by the scaling transformation $\sqrt{\theta} \phi_i \to \sqrt{\theta'} \phi_i$.
Under this transformation, the Lagrangian changes $\mathcal{L}_E[\theta,g] \to (\theta'/\theta) \mathcal{L}_E[\theta',g']$ with $g' \equiv (\theta'/\theta)g$.
Thus the quartic interaction becomes more repulsive when $\theta < \theta'$.
This argument is independent of the sign of $g$.
If attractive interaction $g<0$ were introduced (even though the absolute value of $\phi_i$ is unbound), the strength of the attractive interaction would be enhanced by the statistical parameter.
The dependence on the spatial volume $L$ is shown in Fig.~\ref{figN}.
The results are insensitive to the spatial volume, so that this is not a phase transition.
As the statistical parameter increases, the number density decreases but does not go to zero even in the large volume limit.
These results imply that the statistical parameter causes only quantitative change, not qualitative change, in this setup.

As shown in the previous section, even if we start with the non-interacting anyon field in the original Hamiltonian \eqref{eqHA}, many-body interaction is induced in the Hamiltonian \eqref{eqHB} or the Lagrangian \eqref{eqLB}.
While $\theta$ is originally introduced as the intrinsic constant that is uniquely determined by particle statistics, it can be viewed as the coupling parameter of such statistically induced interaction.
The anyon theory has nontrivial physical properties due to the presence of the statistically induced interaction.
For example, a quantum phase transition can take place along the axis of $\theta$.
Such a statistically induced phase transition was found in non-relativistic theory \cite{2011NatCo...2E.361K} but not known in relativistic theory in 1+1 dimensions.
Although we could not find any tendency of phase transitions in this simulation, we might find it in other parameter regions, e.g., at low temperatures or with other potential terms, by spending more time and effort on parameter search.

Two remarks are in order.
The first one is about the $2\pi$-periodicity of $\theta$.
We have transformed the Hamiltonian to the Lagrangian, and then discretized it on the lattice.
Even if the continuous path integral has the $2\pi$-periodicity, the lattice discretization breaks it.
We numerically checked that the $2\pi$-periodicity is actually broken.
The $2\pi$-periodicity recovers only in the continuum limit.
An alternative way to preserve the $2\pi$-periodicity is the transformation from the Hamiltonian to the Lagrangian after the lattice discretization.
The lattice Hamiltonian is, however, not quadratic in $\pi_i$ but exponential of $\pi_i$.
Thus the integration of $\pi_i$ becomes troublesome.
These are similar to the compact and non-compact formulations of U(1) lattice gauge theory.
The periodicity of the gauge field is present in the compact U(1) gauge theory but absent in the non-compact U(1) gauge theory.
In this sense, the present formulation is the ``non-compact'' one.
It corresponds to the expansion around $\theta=0$, so that the application is limited to small values of $\theta$.
The second one is about the continuum limit.
Since the theory is rewritten by interacting scalar field theory, the continuum limit can be discussed in the usual manner.
At the perturbative level, it can be checked by expanding the lattice action in the powers of $\theta$.
Beyond the perturbation, there are non-trivial questions; how is the renormalization group flow, and whether the compact and non-compact formulations are consistent.

\section{Comments}

We have formulated anyon field theory based on the commutation relation \eqref{eqCCA}.
The commutation relation \eqref{eqCCA} has an ambiguity at $x=y$.
We have taken the bosonic choice $\theta\sgn(0)=0$.
The constructed anyon exhibits bosonic local properties, e.g., superfluidity.
As another choice, we can take the fermionic choice $\theta\sgn(0)=\pi$.
The anyon field is constructed by the Jordan-Wigner transformation of a fermion field.
The four-fermion interaction term, such as $\theta \bar{\psi} \gamma^0 \psi \bar{\psi} \gamma^1 \psi$, is induced.

We would like to close this paper by listing some extensions.
As explained above, the naive circulation of (1+1)-dimensional anyons is trivial.
It would be interesting if we could introduce nontrivial topology by changing space-time geometry or boundary conditions.
The nontrivial topology could also be introduced by the circulation in the space-time plane \cite{Fateev:1985mm}.
If the nontrivial topology exists, the extension to non-Abelian anyons makes sense.
The extension to higher dimensions would be interesting, too.
Since spontaneous symmetry breaking and the condensation are forbidden in 1+1 dimensions, we have studied the behavior of the number density.
In higher dimensions, we can calculate the Bose-Einstein condensate and study its phase transition.

\ack
The author was supported by JSPS KAKENHI (Grant No.~JP15K17624).   
The numerical simulation was carried out on SX-ACE in Osaka University

\bibliographystyle{ptephy}
\bibliography{paper}

\end{document}